\documentclass[prl,showpacs,superscriptaddress,twocolumn]{revtex4-1}
\usepackage{amssymb,amsmath,color,graphicx,psfrag}
\usepackage[normalem]{ulem}

\begin{document}

\newcommand{\kv}[0]{\mathbf{k}}
\newcommand{\Rv}[0]{\mathbf{R}}
\newcommand{\rv}[0]{\mathbf{r}}
\newcommand{\K}[0]{\mathbf{K}}
\newcommand{\Kp}[0]{\mathbf{K'}}
\newcommand{\dkv}[0]{\delta\kv}
\newcommand{\dkx}[0]{\delta k_{x}}
\newcommand{\dky}[0]{\delta k_{y}}
\newcommand{\dk}[0]{\delta k}
\newcommand{\cv}[0]{\mathbf{c}}
\newcommand{\qv}[0]{\mathbf{q}}
\newcommand{\pnt}[1]{\psfrag{#1}{\tiny{$#1$}}}

\newcommand{\sket}[1]{|#1 \rangle_\#} 
\newcommand{\ket}[1]{|#1\rangle} 
\newcommand{\bra}[1]{\langle #1 |} 

\newcommand{\jav}[1]{{\color{red}#1}}

\definecolor{darkgreen}{rgb}{0,0.5,0}
\definecolor{orange}{rgb}{1,0.5,0}
\definecolor{grey}{rgb}{.6,.6,.6}

\newcommand{\scrap}[1]{{\color{orange}{\sout{#1}}}}

\title{Entropy growth and correlation decay in dissipative Luttinger liquid}
\title{Vaporization dynamics of a dissipative quantum liquid}

\author{\'Ad\'am B\'acsi}
\affiliation{MTA-BME Lend\"ulet Topology and Correlation Research Group,
Budapest University of Technology and Economics, 1521 Budapest, Hungary}
\affiliation{Department of Mathematics and Computational Sciences, Sz\'echenyi Istv\'an University, 9026 Gy\H or, Hungary}
\author{C\u at\u alin Pa\c scu Moca}
\affiliation{MTA-BME Quantum Dynamics and Correlations Research Group, Budapest University of Technology and Economics, 1521, Budapest, Hungary}
\affiliation{Department  of  Physics,  University  of  Oradea,  410087,  Oradea,  Romania}
\author{Gergely Zar\'and}
\affiliation{MTA-BME Quantum Dynamics and Correlations Research Group, Budapest University of Technology and Economics, 1521, Budapest, Hungary}
\affiliation{BME-MTA Exotic Quantum Phases Research Group,
Department of Theoretical Physics, Budapest University of Technology and Economics, Budapest, Hungary}
\author{Bal\'azs D\'ora}
\email{dora@eik.bme.hu}
\affiliation{MTA-BME Lend\"ulet Topology and Correlation Research Group,
Budapest University of Technology and Economics, 1521 Budapest, Hungary}
\affiliation{Department of Theoretical Physics, Budapest University of Technology and Economics, Budapest, Hungary}
\date{\today}

\begin{abstract}
We investigate the stability of a
   Luttinger liquid, upon suddenly coupling it to a dissipative environment.
Within the Lindblad equation, the environment couples to local currents and heats the quantum liquid up to infinite temperatures.
The single particle density  matrix
reveals the fractionalization of fermionic excitations in the spatial correlations by retaining the initial non-integer power law exponents, accompanied by
an exponential decay in time with interaction dependent rate.
The spectrum of the time evolved density matrix is gapped, which collapses gradually as $-\ln(t)$.
The von Neumann entropy crosses over from the early time $-t\ln(t)$ behaviour to $\ln(t)$ growth for late times.
The early time dynamics is captured numerically
by performing simulations on spinless interacting fermions, using several numerically exact methods.
Our results could be tested experimentally in bosonic Luttinger liquids.
\end{abstract}

\maketitle

\paragraph{Introduction.}
While dissipation is traditionally viewed as detrimental due to causing decay and randomization of phase,
recent years have witnessed a tremendous progress both in experiment and theory, as a result of which dissipation can now be considered as a useful tool or probe.
Coupling to environment, combined with the ability to create and manipulate  quantum systems\cite{polkovnikovrmp,dziarmagareview,BlochDalibardZwerger_RMP08} in a 
controlled manner, has provided us with unique states of matter\cite{diehl,pichler,barreiro,buca,naghiloo,nhkitaev2018,ashidaprl2,Bardyn2013,Syassen,wineland}, where
dissipation plays a major role.
Such states also hold the promise to be relevant for quantum technologies\cite{reiter}.

Besides  the properties of the steady state, the route towards reaching it  can also reveal a plethora of peculiar phenomena. 
The most prominent example includes quantum effects near the event horizon of a black hole, which  give rise to the 
celebrated Hawking radiation\cite{hawking,parentani} and eventually to black hole evaporation.
In condensed matter and cold atoms context, it is rather  natural to consider the dynamics of 
open quantum systems as these are never perfectly isolated from the environment. Consequently,  
several dissipative many-body systems were investigated\cite{bernier2020,cai,medvedyeva,alba,bernier2018,ashida18,dallatorre}, focusing on 
the propagation and spreading of correlations, quantum information loss, exponential vs. power law temporal relaxation towards 
the steady state as well as the stability of various phases when coupled to a bath\cite{Kosov,fischer}.

Quantum many-body effects are particularly amplified in  one spatial dimension~\cite{giamarchi,nersesyan}. 
In the resulting Luttinger liquid (LL) phase, the original fermionic excitations fractionalize\cite{kamata} into bosonic collective modes due to interactions.
This phase of matter is realized in a variety of fermionic, bosonic, anyonic etc. systems,  including condensed matter\cite{giamarchi} and 
cold atomic systems\cite{cazalillarmp}, quantum optics\cite{changnatphys} and even in black holes\cite{balasll}, and 
promises to be a building block in possible application in topological quantum computation, spintronics and quantum information theory.
This motivated us to combine dissipation with strong correlations and  focus on  the stability and 
evaporation dynamics of LLs by coupling it to a dissipative environment, modeled by the Lindblad equation.
We find that the fermionic single particle density matrix retains its initial LL correlations in space in terms of non-integer power law exponents, 
but the amplitude is reduced in time due to
dephasing.  
This indicates, that  fractionalization persists in spatial correlations.

The von Neumann entropy crosses over from $-t\ln(t)$ for early times to $\ln(t)$ growth for late times.
The early time dynamics is benchmarked numerically with dissipative interacting fermions. Our results are also relevant
for bosonic Luttinger liquids\cite{cazalillarmp}.

\paragraph{Dissipation in the interacting Luttinger model.} 
The low-energy behavior of one-dimensional systems is described by the Luttinger model whose Hamiltonian reads
\begin{gather}
H=\sum_{q>0}\omega_q\left(b^{+}_qb_q + b^{+}_{-q}b_{-q}\right)+g_q\left(b_{q}^+b_{-q}^+ + b_{q}b_{-q}\right)
\label{eq:tdham}
\end{gather}
where $\omega_q=v|q|$, $g_q = g_2|q|$ and $b_q$ annihilates a bosonic excitation.  
Here $v=v_0 + g_4$ is the sound velocity, where $v_0$ is the bare
sound velocity and $g_2$ and $g_4$ describes forward scattering between fermions with different and same chiralities, respectively\cite{giamarchi}.
Since the Hamiltonian is quadratic in the 
bosonic operators, it can be diagonalized by the Bogoliubov transformation, yielding
\begin{gather}
H=E_{GS} + \sum_{q>0}\tilde\omega_q\left(d_q^+d_q + d_{-q}^+d_{-q}\right)
\end{gather}
where $E_{GS}= \sum_{q>0} \left(\tilde\omega_q - \omega_q\right)$ is the ground state energy 
and $\tilde\omega_q=\tilde v|q|$ is the spectrum of elementary excitations with the renormalized sound velocity $\tilde v=\sqrt{v^2-g_2^2}$.

We consider a LL, prepared in the ground state of the interacting Hamiltonian thus no excitations are present. 
At $t=0$, the coupling between the LL and its environment is switched on, and for $t>0$, the time evolution is governed by the Lindblad 
equation\cite{daley,carmichael,breuer}.
The coupling to environment is modeled by local current operators, as in Refs. \cite{eisler2011,temme2012,pereverzev,alba,horstmann}. Such dissipators arise naturally when considering 
fluctuating vector potential or gauge field as the environment.
The Lindblad equation reads as
\begin{gather}
\partial_t \rho = -i[H,\rho] + \gamma\int \mathrm{d}x\, \left([j(x),\rho j(x)] + h. c.\right)
\label{eq:lindbladint0}
\end{gather}
where $\rho(t)$ is the density matrix of the system and $j(x)$ is the current operator playing the role of the jump operator.
Using bosonization \cite{giamarchi}, the current operator is \cite{EPAPS}
\begin{gather}
j(x)= \sum_{q\neq 0}\sqrt{\frac{|q|}{2L\pi}}\textmd{sgn}(q)e^{-iqx}\left(b_{-q}-b^{+}_q\right)
\end{gather}
with $L$ the system size and the spatial integral in Eq. \eqref{eq:lindbladint0} results in
\begin{gather}
\partial_t \rho = -i[H,\rho] + \frac{\gamma}{2\pi}\sum_{q\neq 0} \left([L_q,\rho L_q^+] + h. c.\right)
\label{eq:lindbladint}
\end{gather}
with $L_q=\sqrt{|q|}\left( b_q - b_{-q}^+ \right)$. The spectrum of Eq. \eqref{eq:lindbladint} is expected to be gapless since the 
energy scale in both the Hamiltonian and the dissipator $\sim |q|$.
After Bogoliubov transformation, the jump operator is rewritten as
$L_q =\sqrt{\frac{|q|}{K}}\left(d_q - d_{-q}^{+}\right)$,
where $K=\sqrt{(v-g_2)/(v+g_2)}$ is the Luttinger parameter\cite{giamarchi} and $K<1$ ($K>1$) for repulsive (attractive) interaction.
The presence of the interaction induces a renormalization of the dissipative coupling  $\gamma\rightarrow \gamma/ K$.
This indicates that dissipation becomes effectively stronger/weaker for repulsive/attractive interaction for the density matrix, respectively.

Based on the Lindblad equation, the expectation values of the occupation number and the anomalous operator are obtained as
\begin{subequations}
\begin{gather}
n_q(t)=\textmd{Tr}\left[\rho(t)d_q^{+}d_q\right] = \gamma |q| t/(\pi K)  \label{eq:n}\\
m_q(t)=\textmd{Tr}\left[\rho(t)d_q^{+}d_{-q}^{+}\right] = \frac{\gamma}{2\pi i K\tilde v}\left(e^{i2\tilde v|q| t}-1\right) \label{eq:m}
\end{gather}
\label{eq:nm}
\end{subequations}
in accordance with Ref. \cite{buchhold2014}.
The linear increase of the occupation number implies that the system heats up to infinite temperatures~\footnote{The heating does \emph{not} occur through
thermal density matrices with increasing temperature but rather through highly non-equilibrium non-thermal  density matrices. These are incarnated in the
 non-thermal response of the Green's function.} and the LL eventually evaporates during the Lindblad dynamics, unlike
the related problem with localized loss\cite{froml,dolgirev}.
This is also follows from the observation that the jump operator is hermitian.

\paragraph{Green's function.}
To have a deeper understanding of correlations, we study the time evolution of the single particle density matrix or equal time Green's function defined as
\begin{gather}
G(x,t)=\mathrm{Tr}\left[\rho(t)\Psi_R^+(x)\Psi_R(0) \right]
\label{eq:greendef}
\end{gather}
where 
$\Psi_R(x)=\frac{1}{\sqrt{2\pi\alpha}}
\exp\left[i\sum_{q>0}\sqrt{\frac{2\pi}{qL}}\left(e^{iqx}b_q+e^{-iqx}b^+_q \right)\right]$
is the fermionic field operator of right-moving electrons.
By evaluating the trace in Eq. \eqref{eq:greendef}, the single particle density matrix is obtained as\cite{EPAPS}
\begin{gather}
\ln \frac{G(x,t)}{G_0(x)}=\sum_{q>0}\frac{8\pi}{L|q|}\left(\frac{g_2}{\tilde v}\textmd{Re}\,m_q(t)-\frac{v}{\tilde v}n_q(t)\right)\sin^2\left(\frac{qx}{2}\right)
\label{eq:green4}
\end{gather}
where 
$G_0(x)=\frac{i}{2\pi(x+i\alpha)}\left(\frac{\alpha}{\sqrt{x^2+\alpha^2}}\right)^{\frac{K+K^{-1}}{2}-1}$
is the initial Green's function obeying the well-known \cite{giamarchi} power-law decay for $x\gg\alpha$ with the 
exponent of $(K+K^{-1})/2$ . The momentum summation  is 
regularized with the exponential cutoff $\exp(-\alpha |q|)$ with $\alpha$ the short distance cutoff.

\begin{figure}[t!]
\centering
\includegraphics[width=6cm]{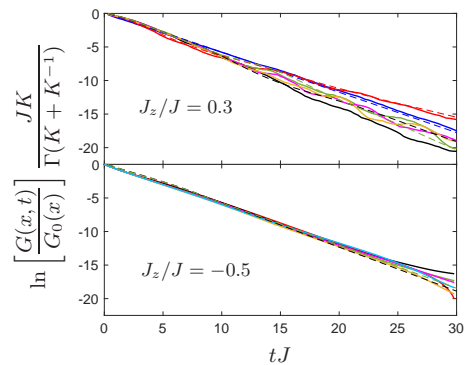}
\caption{The early time scaling of the Green's function for various $x$ values, obtained
 using three distinct numerical methods. The Green's function decays with the same interaction dependent exponent at each spatial separation, $x$. 
Top panel: $J_z/J=0.3$, $\Gamma/J=0.04$  and 
 $N=22$ (thick solid line)
using the quantum jump method with ED and PBC and 6000 averages over quantum trajectories and for
$N=14$ (thin dashed line) using  ED with PBC for the Lindblad equation.
Bottom panel: $J_z/J=-0.5$, $\Gamma/J=0.4$  and
 $N=41$ using TDVP (thick solid line) with OBC and for $N=14$
(thin dashed line) using  ED with PBC for the Lindblad equation.
The agreement between various methods indicate that the data is relatively free 
from finite size effects.
Here, $x$=1, 3, 5, 7, 9, 11, 13 (blue, red, black, green, magenta, gold and light blue, respectively), but not all $x$'s are shown.}
\label{fig:green}
\end{figure}

It is important to note that the time-dependence of the single particle density matrix occurs only 
through the quantities $n_q(t)$ and $m_q(t)$ which have been calculated in Eqs. \eqref{eq:nm}. Substituting these into Eq. \eqref{eq:green4}, the summation over $q$ is carried out analytically as
\begin{gather}
\ln \frac{G(x,t)}{G_0(x)}=-\frac{\gamma t}{\pi\alpha}\frac{K^{-2}+1}{\left(\frac{\alpha}{x}\right)^2+1} +\frac{\gamma}{2\pi \tilde v}\left(\frac{1}{K^2}-1 \right)I
\left(\frac{\tilde vt}{\alpha},\frac{x}{\alpha}\right)
\label{eq:green6}
\end{gather}
where
$I(y,z) = \arctan(2y) - \sum\limits_{s=\pm}\frac{\arctan(2y-sz)}{2}$.
In the scaling limit, when $(x,\tilde v t)\gg\alpha$, the time evolution of the single particle density matrix is 
summarized as
\begin{gather}
G(x,t)=\frac{i}{2\pi\alpha}\left(\frac{\alpha}{x}\right)^{\frac{K+K^{-1}}{2}}\exp\left(-\frac{(K^{-1}+K)\gamma t}{\pi\alpha K}\right)
\times \nonumber \\ 
\times\left\{\begin{array}{cc}
\exp\left(\frac{\gamma}{4\tilde v}(K^{-2}-1)\right) & \textmd{ for } 2 \tilde v t\ll x\\
1 & \textmd{ for }2\tilde v t\gg x
\end{array}\right.
\label{greenlimits}.
\end{gather}
It exhibits two peculiar phenomena: the power law spatial decay of the single particle density matrix is preserved throughout the time evolution
with the  initial LL exponent of $(K+K^{-1})/2$.
This non-integer exponent indicates that part of the original fermionic excitations remain
fractionalized during the non-unitary time evolution.  
In addition, the spatial correlations are uniformly suppressed, exponentially in time, in accord with Ref. \cite{eisler2011}.
The characteristic time scale of the dephasing is set by the dissipative coupling and the interaction strength as $K\pi\alpha/(\gamma(K+K^{-1}))$, as found numerically 
in Fig. \ref{fig:green}. The decay rate decreases from attractive ($K>1$) to repulsive ($K<1$) interaction: even though $\gamma$ itself is renormalized to $\gamma/K$ 
in the Lindblad equation, the original bare fermion, $\Psi_R(x)$ is also dressed by the interaction, thus reverting the trend for the Green's function. 
It is rather remarkable that in spite of the gapless spectrum of the Lindbladian\cite{znidaric},
the fermionic Green's function still decays exponentially in time.
On top of this, one may observe a kink in the single particle density matrix which travels with the velocity $2\tilde v$, which
is the only light-cone effect, though this
is rather minor and is expected to be hardly observable. The behaviour in Eq. \eqref{greenlimits} 
is rather generic and occurs for other correlation functions as well\cite{EPAPS}.

\begin{figure}[t!]
\centering
\includegraphics[width=5.5cm]{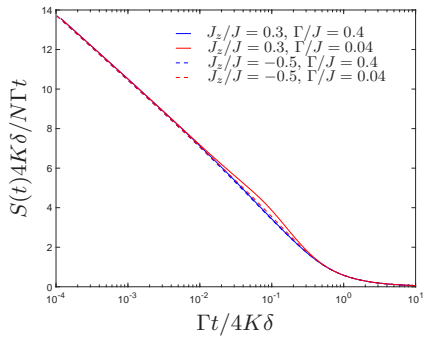}
\caption{The early time scaling of the von Neumann entropy is shown for $N=14$ using ED for various parameters. Hardly any 
finite size effects are present since the $N=10$
data falls almost on top of this. The parameter $\delta=\alpha(J_z)/\alpha(0)$ accounts for the renormalization of
 $\alpha$ with interaction, and is expected to increase\cite{pollmannxxz} with $J_z$. Here we used
$\delta=0.73$ and $1.15$ for $J_z/J=-0.5$ and 0.3, respectively with $\delta=1$ for the non-interacting case. }
\label{fig:entropy}
\end{figure}

\paragraph{Time evolved density matrix and entropy.}
Another interesting quantity which characterizes the time evolution governed by the Lindblad equation,
 is the von Neumann or thermodynamic entropy defined as $S(t)=-\textmd{Tr}\left[\rho(t)\ln\rho(t)\right]$.
With the bosonized version of $\rho(t)$\cite{EPAPS}, the trace is evaluated as
\begin{gather}
S(t)=2\sum_{q>0}\left[(N_q(t)+1)\ln(N_q(t)+1) - N_q(t)\ln N_q(t)\right],
\label{eq:entropy}
\end{gather}
where
$N_q(t)=\sqrt{\left(n_q(t)+\frac{1}{2}\right)^2-|m_q(t)|^2}-\frac{1}{2}$.
 Interestingly, the time-dependence occurs again only through the functions given in Eq. \eqref{eq:nm}.
Its  early and long time limits are calculated as
\begin{gather}
S(t)\sim \frac{L}{\pi\alpha}
\left\{\begin{array}{cc} -\dfrac{\gamma t}{K\pi\alpha} \ln\left(\dfrac{\gamma t}{K\pi\alpha}\right)  & \textmd{for $\gamma t \ll K\pi\alpha$} \\ 
\ln\left( \dfrac{\gamma t}{K\pi\alpha}\right)  & \textmd{for $\gamma t \gg K\pi\alpha$} \end{array}\right.
\label{entgrowth}
\end{gather}
The early time growth agrees with numerics on dissipative interacting fermions in Fig. \ref{fig:entropy}, while the 
latter\footnote{The $\ln(t)$ late time entropy growth is analogous to the high
temperature ($T$) equilibrium entropy of one dimensional acoustic phonons $\sim\ln(T)$ for temperatures  much larger than the bandwidth.} is reminiscent of
the behaviour of the entanglement entropy in many-body localized systems\cite{znidaric2008,pollmann2012}.

In order to understand more closely the origin of this behaviour, we can evaluate also the eigenvalues of the time evolved density matrix at each time instant, denoted
by $\lambda_0\geq\lambda_1\geq\lambda_2\dots$. Formally, we can also assign an instantaneous Hamiltonian to the time evolved density matrix, $\rho(t)=\exp(-H_\rho(t))$,
whose spectrum is $-\ln\lambda_i$. We can define the gap in the many-body spectrum  as $\Delta_\rho=\ln(\lambda_0/\lambda_1)$.
This is analogous to the spectrum of the reduced density matrix and the corresponding entanglement Hamiltonian and entanglement gap
in closed quantum systems\cite{thomale2010,chandran}. Since the initial state is pure, the $t=0$ 
spectrum is trivial\footnote{There is one 1 eigenvalue of the initial density matrix, while all the others are zero. This translates into a infinitely 
large gap in the spectrum of the initial $H_\rho(t=0)$.}. During the time evolution, the density matrix is brought to diagonal form after an instantaneous Bogoliubov
transformation as $\rho(t)\sim\exp(-\sum_q\Omega_q(t)\tilde b^+_q\tilde b_q)$, and for each momentum sector, the single particle spectrum is 
$\Omega_q(t)=\ln\left(1+\frac{1}{N_q(t)}\right)$. At $t=0$, all $N_q(t=0)=0$, therefore $\Omega_q(t=0)=\infty$, and the $\tilde b_q$ bosons are in their vacuum state,
 the gap in the spectrum is infinitely large.
After switching on the dissipation, the gap in the many-body spectrum,
which parallels closely to the entanglement gap, starts to decrease slowly for early times as
\begin{gather}
\Delta_\rho\approx\ln\left(\frac{\pi K\alpha}{\gamma t}\right).
\label{entgap}
\end{gather}
The bosonization approach is valid for momenta $|q|<1/\alpha$. Our analytical results
show that these modes definitely give a singular, $t\ln(t)$ and $\ln(t)$ contribution to the entropy and to the
gap in the many-body spectrum
at short
times, respectively. We cannot determine analytically the contribution of the high energy modes,
which lie outside the range of the bosonization approach. However, our numerics
is indicative that the contribution of these high energy modes is subleading,
 compared to the LL contribution.

\paragraph{Interacting fermions within the Lindblad equation.}

To illustrate our findings and check their validity in lattice models, 
we have investigated one dimensional spinless fermions in an \emph{open} 
tight-binding chain with nearest neighbour interaction at half filling
using several numerical techniques. The closed system is
equivalent to the 1D Heisenberg XXZ chain after a Jordan-Wigner transformation\cite{giamarchi,nersesyan}.
The Hamiltonian is
\begin{gather}
H=\sum_{m=1}^{N}\left[ \frac{J}{2} \left(c^+_{m+1}c_m +c^+_{m}c_{m+1}\right)+J_z n_{m+1}n_m\right],
\label{xxz}
\end{gather}
where $c$'s are fermionic operators, $n_m=c^+_{m}c_m$ and $J_z$ denotes the nearest neighbour repulsion, $N$
the number of lattice sites and the model hosts $N/2$ fermions.  This model realizes a LL for $|J_z|<J$
and the strength of the interaction is characterized by the dimensionless  LL parameter\cite{giamarchi} $K=\pi/2[\pi-\arccos(J_z/J)]$ from the Bethe Ansatz solution
of the model. Due to the bounded spectrum of Eq. \eqref{xxz}, the bosonization results are only applicable for early times, before
the whole band is populated during heating.


The lattice version of the current operator in Eq. \eqref{eq:lindbladint0} reads as
\begin{gather}
j_m=i\left(c^+_{m+1}c_m -c^+_{m}c_{m+1}\right)/2,
\end{gather}
which appears in the environmental part of the Lindblad equation as $\Gamma\sum_m \left([j_m,\rho j_m] + h. c.\right)$. 
To make contact with bosonization, we use $\gamma/\alpha\sim\Gamma$.
A similar problem with different jump operator\footnote{The local 
current is more non-local than the local density: the local densities as jump operators yield presumably simpler
dynamics, as these operators commute with each other (unlike the local currents), and arbitrary power of the local density equals to the local density itself.}
was considered in Refs. \cite{bernier2020,cai,medvedyeva}. 

The Lindblad equation for this dissipative many-body system is attacked by three different methods.
By vec-ing\cite{shallem}, i.e. rearranging the square density matrix as a vector, one can use standard exact diagonalization (ED) and Krylov-space
time evolution, reaching $N=14$. Second, using the quantum jump method\cite{daley,pichler,carmichael} for the same system, we can reach $N=22$ at 
the expense of having to average over
the quantum trajectories. For these two methods, periodic boundary condition (PBC) is used to minimize finite size effects.
Finally, we use the time dependent variational principle (TDVP) with open boundary condition (OBC)\cite{Haegeman.2011, Haegeman.2013,SciPostPhysLectNotes.7}
within the matrix product states  framework, to directly simulate the density matrix.
Initially, we prepare the system in the ground state by using the
density matrix renormalization group\cite{White-1992}, and use the ground state $\ket{\Psi_0}$ to
build the density matrix $\rho_0=\ket{\Psi_0}\bra{\Psi_0}$ in the form of a
matrix product  operator.
Next, by vec-ing the density matrix to $\sket{\rho}$
the Lindblad equation~\eqref{eq:lindbladint} is rewritten as
$\partial_t \sket{\rho (t)}={\cal L} \sket{\rho(t)} $, with ${\cal L} $ the Lindbladian organized now as a matrix product  operator.

Using these techniques, we determine the equal time Green's function, i.e. 
$G(x,t)=\textmd{Tr}\left(\rho(t)c^+_{m+x}c_{m}\right)$. For PBC,  this becomes independent of $m$
due to translational invariance, while for OBC, $m$ and $m+x$ are chosen symmetrically to the chain center to reduce the effects from boundary condition. As expected,
$G(0,t)=1/2$ is recovered in all numerics (not shown). The spatio-temporal dynamics of the single particle density matrix is  
plotted in Fig. \ref{fig:green}, confirming the results
of bosonization: the spatial and temporal dynamics practically decouples, the former preserves the LL correlation encoded in the initial state, while the latter
displays pure dephasing for short times, analogously to Ref. \cite{eisler2011}. 
 However, the temporal decay rate is strongly influenced by the LL parameter $K$, and decreases monotonically with the interaction. 
The curves for different $J_z$'s are not a priori expected to fall on top of each other as $\alpha$ in Eq. \eqref{eq:green6} can follow a weak $J_z$ dependence.
For longer times,
deviations from the bosonization results are expected when the explicit nature of the high energy degrees come into play. These induce
 model dependent\cite{bernier2020}, 
non-universal features, whose study is beyond the scope of our current work.

\begin{figure}[h!]
\centering
\includegraphics[width=6cm]{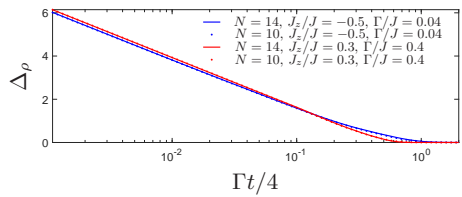}
\caption{The early time scaling of the gap in the spectrum of the instantaneous Hamiltonian of the time evolved density matrix is plotted for several parameters.
It agrees with Eq. \eqref{entgap} and is free from finite size effects.}
\label{fig:entanglement}
\end{figure}

With the knowledge of the time dependent density matrix, the dynamics of the von Neumann entropy is evaluated.
For early times, it follows the expected $-\Gamma t\ln(\Gamma t)$ early time growth, and obeys the scaling form predicted by bosonization, as shown 
in Fig. \ref{fig:entropy}. 
Here we had to account for the mild interaction dependence of the cutoff by slightly renormalizing 
the value of the rate $\Gamma\rightarrow \Gamma/\delta$\cite{pollmannxxz}. 
Distinct
cutoff dependent physical quantities, i.e. the single particle density matrix vs. entropy, 
 may require slightly different interaction dependence of the cutoff.
The explicit value of the decay rate for a given microscopic model 
can be determined similarly to the gap in sine-Gordon related models\cite{giamarchi} by comparing the analytical results to numerics for the time dependent 
entropy and correlation functions.
For late times, the entropy converges fast to its maximal value on the lattice $\sim N\ln(2)$ 
and the $\ln(t)$ late time growth 
of the LL is not reproduced due to the small local Hilbert space dimension (i.e. 2) for fermions.
We speculate that this  late time growth could possibly show up in bosonic realization of LLs\cite{cazalillarmp}, where
the local Hilbert space is much bigger\footnote{For low fillings $1/k$ with $k\gg 1$, the maximal 
entropy $\sim N_p\ln(k)$ with $N_p$ the total number of bosons. For small enough $1/k$, there is enough room for the $\ln(t)$ growth to develop before saturating
to the maximal value.}.

Finally, we evaluate the gap in the spectrum of the time evolved density matrix, as discussed above. Its numerically obtained value is shown in Fig. \ref{fig:entanglement}, 
which,
in spite of its  cutoff dependence, still follows the $-\ln(\Gamma t)$ prediction of bosonization.

\paragraph{Summary.}
We have studied the vaporization dynamics of Luttinger liquids after coupling to to dissipative environment through the local currents.
Unlike unitary quantum quenches, where the dynamical Luttinger liquid exponents are different from the equilibrium
ones\cite{cazalillaprl}, in our case the single particle density matrix reveals the persistence
of fractionalization of fermionic excitations in  spatial correlations with the equilibrium exponents,
but with an amplitude exponentially suppressed in time.

The von Neumann entropy crosses over from an early time $-t\ln(t)$ growth to $\ln(t)$ growth for late times.
The former is attributed to the logarithmic in time collapse of the instantaneous gap in the time evolved density matrix.
The early time features are captured numerically in a dissipative interacting fermionic lattice model.
Our results apply to a large variety of systems and are observable in bosonic Luttinger liquids.

\begin{acknowledgments}
This research is supported by the National Research, Development and Innovation Office - NKFIH   within the Quantum Technology National Excellence Program (Project No.
      2017-1.2.1-NKP-2017-00001), K119442, K134437, SNN118028 and by the BME-Nanotechnology FIKP grant (BME FIKP-NAT).
\end{acknowledgments}

\bibliographystyle{apsrev}
\bibliography{lindblad,wboson1}

\pagebreak

\section{Supplementary material for "Vaporization dynamics of a dissipative quantum liquid"}

\setcounter{equation}{0}
\renewcommand{\theequation}{S\arabic{equation}}

\setcounter{figure}{0}
\renewcommand{\thefigure}{S\arabic{figure}}

\section{The current operator}

The physical fermion field is decomposed into right and left moving excitations as
$\Psi(x)=e^{ik_F x}\Psi_R(x) +e^{-ik_F x}\Psi_L(x)$ with $k_F$ the Fermi wavenumber.
The current operator can be rewritten as $j(x)=j_0+\exp(i2k_Fx)j_{2k_F}+\exp(-i2k_Fx)j_{2k_F}^+$, where the 
 long ($j_0$) and short ($j_{2k_F}$) wavelength current operators are determined from Eq. (15) in the main text in the continuum limit as
\begin{gather}
j_0(x)\sim \Psi_R^+(x)\Psi_R(x)-\Psi_L^+(x)\Psi_L(x),\\
j_{2k_F}\sim \Psi_R^+(x)\partial_x\Psi_L(x)-\left[\partial_x\Psi_R^+(x)\right]\Psi_L(x).
\end{gather}
Notably, the second expression contains an additional gradient compared to the first one, which increases its scaling dimension by one and  is
 considered to be more irrelevant than the long wavelength, $q\sim 0$
 term in equilibrium.
We assume that this classification remains also valid for the early time dynamics of Lindblad description, and therefore retain only $j_0$ in the jump operator.
This is verified by comparing bosonization to numerics.

\section{Time evolution of the single particle density matrix}
The single particle density matrix is defined in Eq. (7) of the main text. 
Following standard steps \cite{giamarchi,cazalillaprl}, we obtain
\begin{gather}
\frac{G(x,t)}{G_\mathrm{nonint}(x)}=
\exp\left(-\sum_{q>0}\frac{4\pi}{Lq}n^B_q(t)\left(1-\cos(qx)\right)\right),
\label{eq:green3}
\end{gather}
where
\begin{gather}
G_\mathrm{nonint}(x)=\frac{i}{2\pi(x+i \alpha)}
\end{gather}
is the non-interacting single particle density matrix and
\begin{gather}
n^{B}_q(t) =\mathrm{Tr}\left[\rho(t)b_q^{+}b_q\right]
\end{gather}
is the instantaneous number of $b$-bosons which describe the elementary excitations of the non-interacting system. After Bogoliubov transformation, the number of $b$-bosons is expressed as
\begin{gather}
n^{B}_q(t) = \frac{v}{\tilde v}n_q(t) - \frac{g_2}{\tilde v}\mathrm{Re}\,m_q(t) + \frac{1}{2}\left(\frac{v}{\tilde v}-1\right)
\label{eq:nB}
\end{gather}
where $n_q(t) = \mathrm{Tr}\left[\rho(t)d_q^+d_q\right]$ and $m_q(t) =\mathrm{Tr}\left[\rho(t)d_q^+d_{-q}^+\right]$. After substituting $n_q^B(t)$ into Eq. \eqref{eq:green3}, the integral of the term with $\frac{1}{2}\left(\frac{v}{\tilde v}-1\right)$ leads to a power-law function of $x$. This function (together with $G_\mathrm{nonint}(x)$) results in the interacting correlation function
\begin{gather}
G_0(x)=\frac{i}{2\pi(x+i\alpha)}\left(\frac{\alpha}{\sqrt{x^2+\alpha^2}}\right)^{\frac{K+K^{-1}}{2}-1}
\end{gather}
which also equals the single particle density matrix in the initial state. The time dependence is described in the first two terms of 
Eq. \eqref{eq:nB} which, after all, end up in Eq. (8) of the main text.

\section{Time evolution of entropy}
In this section, the time dependence of the von Neumann entropy is studied. At any time instant, the system consists of two Bose gases for each $q>0$ quantum numbers. 
Therefore, the entropy is defined  as
\begin{gather}
S(t) = -\mathrm{Tr}\left[\rho(t)\ln\rho(t)\right]=\nonumber \\
= 2\sum_{q>0}\left[(N_q(t)+1)\ln(N_q(t)+1) - N_q(t)\ln N_q(t)\right]
\end{gather}
where $N_q(t) = \mathrm{Tr}\left[\rho(t)\tilde{b}_q^+(t)\tilde{b}_q(t)\right]$ is the number of bosons $\tilde{b}$ which diagonalize the instantaneous density matrix.
To calculate $N_q(t)$ and the entropy, we determine how the operators $\tilde{b}_q(t)$ are related to the operators $d_q$ which diagonalize the interacting Hamiltonian.

In terms of the operators $d_q$, the density matrix is expressed as
\begin{gather}
\rho(t)=\prod_{q>0}r_q(t)e^{c_q(t)K_{q,-}}e^{-2\ln(\nu_q(t)+1)K_{q,0}} e^{c_q(t)^*K_{q,+}}
\label{eq:rhodef}
\end{gather}
where the operators $K_{q,+}=K_{q,-}^{+} = d_q^+d_{-q}^+$ and $K_{q,0} =\frac{d_q^+d_q + d_{-q}d_{-q}^+}{2}$ obey the commutation relations of an $su(1,1)$ algebra.
Note that all the time dependence is incorporated into the functions $c_q(t)$ and $\nu_q(t)$. The prefactor is set to 
$r_q(t)=(\nu_q(t)^2-|c_q(t)|^2)/(\nu_q(t)+1)$ in order to ensure the unit trace in each wavenumber channel. It can be shown that the functions are related to the expectation values $n_q(t)$ and $m_q(t)$, which are obtained in Eqs. (6) of the main text, by
\begin{gather}
\nu_q(t) = \frac{n_q(t)}{n_q(t)^2-|m_q(t)|^2}\quad c_q(t) = \frac{m_q(t)}{n_q(t)^2-|m_q(t)|^2}\,.
\label{eq:nuc}
\end{gather}

To diagonalize the exponent of \eqref{eq:rhodef}, first we rewrite the product of the three exponentials in a single exponential by using the commutation rules of the $su(1,1)$ algebra \cite{solomon,gilmore}.
\begin{gather}
\rho(t)=\prod_{q>0}r_q(t)\,e^{\frac{\Omega_q}{\sqrt{1-|s_q|^2}}\left(s_q K_{q,-}+2K_{q,0} + s_q^*K_{q,+}\right)}
\end{gather}
where
\begin{gather}
\Omega_q = \left|\textmd{acosh}\left(\frac{1}{2(n_q(t)+n_q(t)^2-|m_q(t)|^2)}+1\right)\right|
\label{eq:omegatdef}
\end{gather}
and 
\begin{gather}
s_q = -\frac{2m_q(t)}{1+2n_q(t)}
\label{eq:sdef}
\end{gather}
are both time dependent.
Since the exponent of the density matrix is quadratic in the bosonic annihilation and creation operators, it can be diagonalized by the Bogoliubov transformation
\begin{gather}
\left[\begin{array}{c} \tilde{b}_{q}(t) \\ \tilde{b}_{-q}(t)^{+}\end{array}\right] = \left[\begin{array}{cc} u_q(t) & v_q(t) \\ v_q(t)^{*} & u_q(t) \end{array}\right]
\left[\begin{array}{c} d_{q} \\ d_{-q}^{+}\end{array}\right]
\label{eq:bogoliubov}
\end{gather}
where
\begin{gather}
u_q(t)=\frac{1}{\sqrt{2}}\sqrt{\frac{1}{\sqrt{1-|s_q(t)|^2}}+1} \\
v_q(t)=\frac{s_q(t)^{*}}{\sqrt{2}|s_q(t)|}\sqrt{\frac{1}{\sqrt{1-|s_q(t)|^2}}-1}
\end{gather}
leading to $\rho(t)\sim e^{-\Omega_q(t)\left( \tilde{b}_{q}(t)^+ \tilde{b}_{q}(t) + \tilde{b}_{-q}(t)^{+}\tilde{b}_{-q}(t)\right)}$. For the entropy, we have to calculate the expectation value of the number of bosons $\tilde{b}$. Substituting the Bogoliubov coefficients, we obtain
\begin{gather}
N_q(t) = \mathrm{Tr}\left[\rho(t)\tilde{b}_q^+(t)\tilde{b}_q(t)\right] = \nonumber \\
= (u_q^2+|v_q|^2)n_q(t) + 2\mathrm{Re}\left(u_q v_q m_q(t)\right) + |v_q|^2 = \nonumber \\
= \sqrt{\left(n_q(t) + \frac{1}{2}\right)^2-|m_q(t)|^2}\,\,-\frac{1}{2}
\end{gather}

\section{Spin-flip correlation function}

Eq. (14) in the main text is equivalent to the Heisenberg XXZ chain and can also be rewritten in terms of hard core bosons\cite{giamarchi}.
Then, the hard core boson equals time Green's function or the spin flip correlation function\cite{pollmannxxz,giamarchi} is
\begin{gather}
C(x,t)=\frac{(-1)^x}{2\pi\alpha}\mathrm{Tr}\left[\rho(t)e^{-i\Theta(x)}e^{i\Theta(0)}\right]
\end{gather}
where 
\begin{gather}
\Theta(x) = i\sum_{q\neq 0}\sqrt{\frac{\pi}{2L|q|}}e^{iqx}\left(b_q^+ - b_{-q} \right)
\end{gather}
and the density matrix is given by Eqs. \eqref{eq:rhodef} and \eqref{eq:nuc}. Evaluating the trace, we obtain
\begin{gather}
\ln\frac{C(x,t)}{C_0(x)}=-\sum_{q>0}\frac{4\pi\sin^2\left(\frac{qx}{2}\right)}{LKq}\left(n_q(t) + \mathrm{Re}\,m_q(t)\right)
\end{gather}
where
\begin{gather}
C_0(x)=
\frac{(-1)^x}{2\pi\alpha}\left(\frac{\alpha}{\sqrt{x^2+\alpha^2}}\right)^{\frac{1}{2K}}
\end{gather}
is the initial correlation function.

Using the results in Eqs. (6) of the main text, the sum over wavenumbers can be carried out analytically as
\begin{gather}
\ln\frac{C(x,t)}{C_0(x)}=-\frac{\gamma}{2\pi K^2\tilde v}\left(\frac{\alpha 2\tilde v t}{\alpha^2+x^2}+I\left(\frac{\tilde v t}{\alpha},\frac{x}{\alpha}\right)\right)
\end{gather}
where $I(y,z)$ is defined after Eq. (9) in the main text.
In the scaling limit, i.e., when $2\tilde v t\gg \alpha$ and $x\gg\alpha$, 
\begin{gather}
C(x,t)=C_0(x)e^{-\frac{\gamma t}{\pi K^2 \alpha}}\left\{\begin{array}{lc}
e^{-\frac{\gamma}{4K^2\tilde v }} & \textmd{ for }2\tilde v t\ll x\\
1 & \textmd{ for } x\ll 2\tilde v t
\end{array}\right.
\end{gather}
decays exponentially with time.

\end{document}